# Time-Dependent Oxidative Degradation of WSe$_2$ Nanosheets and Its Influence on HER Catalysis

*Panwad Chavalekvirat[a,b], Wisit Hirunpinyopas[c], and Pawin Iamprasertkun[a,b,\*]*

[a]School of Bio-Chemical Engineering and Technology, Sirindhron International Institute of Technology, Thammasat University, Pathum Thani 12120, Thailand (pawin@siit.tu.ac.th).

[b]Research Unit in Sustainable Electrochemical Intelligent, Thammasat University, Pathum Thani 12120, Thailand.

[c]Department of Chemistry, Faculty of Science, Kasetsart University, Bangkok 10900, Thailand

*Corresponding author: Pawin Iamprasertkun (pawin@siit.tu.ac.th)

https://orcid.org/0000-0001-8950-3330

ABSTRACT

The selection of an appropriate solvent has long been a critical factor in the liquid phase exfoliation (LPE) of two-dimensional transition metal dichalcogenide (2D TMD) nanosheets. N-methyl-2-pyrrolidone (NMP) remains one of the most commonly used solvents due to its favourable surface tension compatibility, particularly with tungsten diselenide ($WSe_2$), which facilitates a relatively high yield of exfoliated nanosheets. However, prolonged exposure of the nanosheets to NMP promotes oxidation, consequently diminishing their hydrogen evolution reaction (HER) activity. In contrast, a more environmentally friendly solvent system, consisting of isopropyl alcohol (IPA) and deionised water (DI), induces significantly less oxidation and results in enhanced HER performance. Notably, this green solvent system achieves the lowest overpotential of 0.209 V vs RHE after 24 hours of chronoamperometry.

KEYWORDS: Liquid Phase Exfoliation, Solvent Selection, $WSe_2$, Electrocatalysts, HER

INTRODUCTION

Transition metal dichalcogenides (TMDs) are promising alternatives to platinum-based catalysts for the hydrogen evolution reaction (HER)[1, 2]. Among them, tungsten diselenide ($WSe_2$) has a suitable band gap, strong photoluminescence, and mechanical strength, contributing to its relatively high HER activity[3-6]. Notably, its electrochemical performance improves significantly after electrochemical cycling[7], with the 2D form showing much higher HER activity than its bulk counterpart[8]. Therefore, synthesising 2D $WSe_2$ nanosheets is essential, and solvent-assisted liquid-phase exfoliation (LPE) provides a simple, cost-effective method for their production[9]. Several LPE parameters affect both exfoliation efficiency and nanosheet quality. One study employed cascade centrifugation and found that $WSe_2$ nanosheets with smaller dimensions have superior HER activity[10]. Exfoliation power also affects the nanosheets' structure and HER performance. Our previous study found that lower power yields $WSe_2$ nanosheets with less oxidation and higher crystallinity, enhancing HER activity[11]. However, oxide content remained around 50% due to the use of NMP, which promotes $WSe_2$ oxidation to $WO_3$, emphasising the importance of solvent selection in optimising nanosheets quality and performance. In this study, we propose the synthesis of $WSe_2$ nanosheets using IPA/DI as a greener alternative to NMP, based on the hypothesis that this solvent mixture induces less oxidation. We also hypothesise that oxidation progresses over time. To evaluate this, we investigate the time-dependent oxidation of $WSe_2$ in both solvent systems over durations ranging from one day to one year.

RESULTS & DISCUSSIONS

The detailed synthesis and characterisation procedures are described in the electronic supplementary information (ESI). As shown in the XRD patterns in Figure 1a, only the WSe$_2$ nanosheets exfoliated in NMP and stored for one year exhibited characteristic peaks of hexagonal WO$_3$ at around 23.6° and 29.8°. Meanwhile, all samples displayed the characteristic peak of 2H-WSe$_2$ at around 13.6°, consistent with previously reported values[12,13]. While XRD provides structural information on WSe$_2$ nanosheets in the solid state, it does not capture local structure changes in dispersion. Therefore, liquid-phase XAS was conducted for a more detailed analysis. The XANES spectra for all WSe$_2$ samples are presented in Figure S1 and Table S1. As shown in Figure 1b, the Fourier transform curves of EXAFS spectra reveal a first-shell W–O peak at approximately 1.5 Å and a W–Se peak near 2.1 Å across all samples. The W–Se peak progressively shifts to higher radial distances and becomes broader with increased time in the solvent, particularly in samples exfoliated in NMP. Meanwhile, the W–O peaks become more pronounced over time, dominating in NMP-exfoliated samples as early as one month; by one year, the W–Se peak is barely distinguishable. In contrast, IPA/DI-exfoliated samples show a relatively balanced presence of W–O and W–Se peaks up to one month, with the W–O peak only becoming dominant after one year (for further details, see Figure S2 and Table S2). These findings indicate time- and solvent-dependent oxidation, especially in NMP, leading to oxidation-induced degradation characterised by increased structural disorder (amorphisation) and the formation of selenium vacancies in the exfoliated WSe$_2$ nanosheets. These results are consistent with previous studies[11,14]. Notably, as XAS was conducted in dispersion, the prominent W–O peaks may partly result from solvent-derived oxygen interacting with the WSe$_2$ nanosheets. The XAS analysis of the solid-state samples is presented in Figure S3.

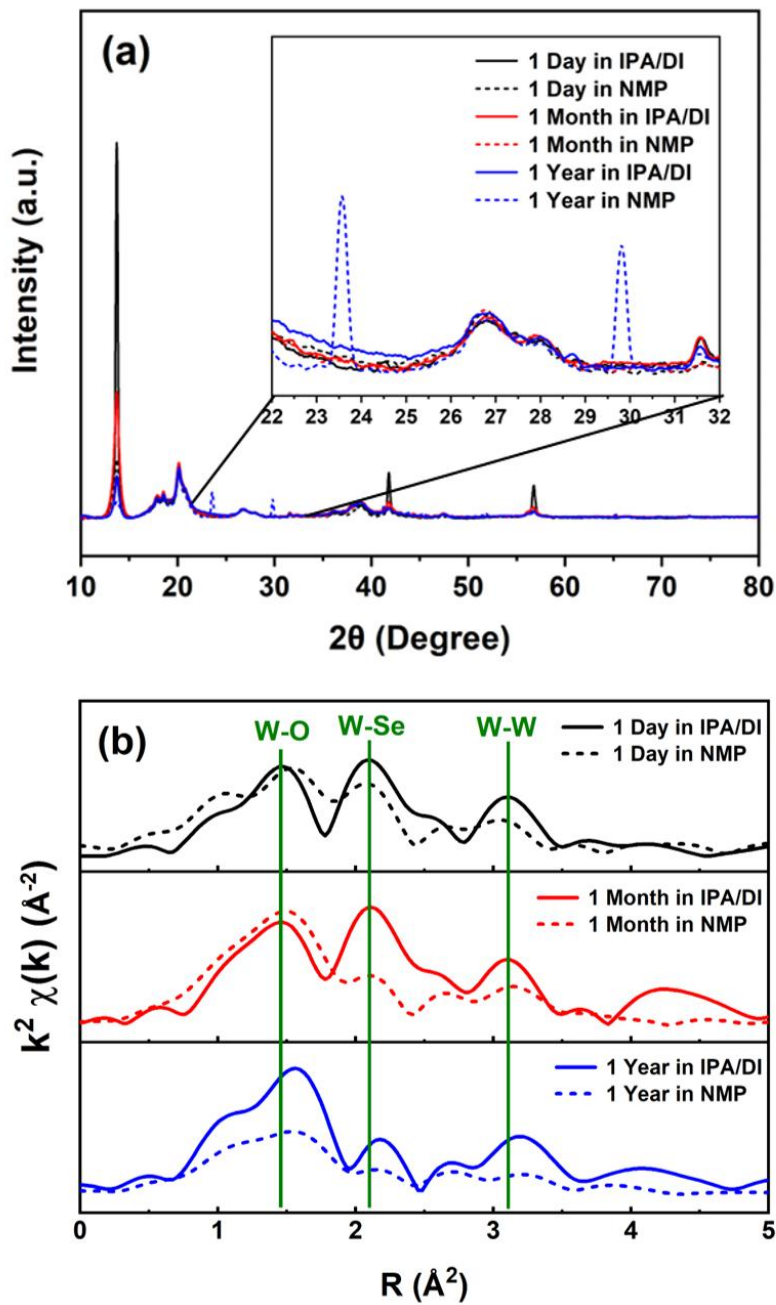

**Figure 1.** (a) X-ray diffraction patterns and (b) radial distances profiles EXAFS spectra of as-exfoliated WSe$_2$ samples in IPA/DI and NMP, for durations ranging from 1 day to 1 year.

To complement the analysis of the local structure, XPS was utilised to investigate the surface chemical composition of the exfoliated $WSe_2$ nanosheets. Figure 2a displays the W 4f XPS spectra for all exfoliated samples, revealing four distinct peaks: the W $4f_{7/2}$ and W $4f_{5/2}$ peaks of the W–Se bond at approximately 32.3 eV and 34.4 eV, as well as the W $4f_{7/2}$ and W $4f_{5/2}$ peaks of the W–O bond at around 35.8 eV and 37.9 eV, consistent with previous reports[15]. The IPA/DI-exfoliated samples exhibit more prominent W–Se peaks compared to those exfoliated in NMP. Over time, the W–O peaks become increasingly pronounced, ultimately dominating the spectra after one year of storage. This trend aligns with the XAS results, confirming the time- and solvent-dependent oxidation behaviour. As shown in Figure 2b, the atomic concentrations of $WSe_2$ and $WO_3$ were quantified. The content of $WO_3$ is consistently higher in NMP-exfoliated samples and increases with storage duration. The IPA/DI-exfoliated sample stored for one day retains the highest $WSe_2$ concentration (~80%), while the NMP-exfoliated sample stored for one year shows a significant decrease to approximately 20%. This accelerated oxidation is attributed to the auto-oxidation of NMP in the presence of oxygen and moisture[16]. This reaction lowers the pH of the medium and facilitates the dissolution of transition metals, leading to the release of free tungsten atoms into the solvent[17]. These free W atoms are then exposed to dissolved oxygen, triggering oxide formation. Moreover, the reactive species produced during NMP auto-oxidation can directly oxidise the $WSe_2$ nanosheets, further promoting the generation of oxide species[16, 18]. Additional XPS data for the Se 3d region and a detailed W 4f peak analysis are included in Figure S4 and Table S3.

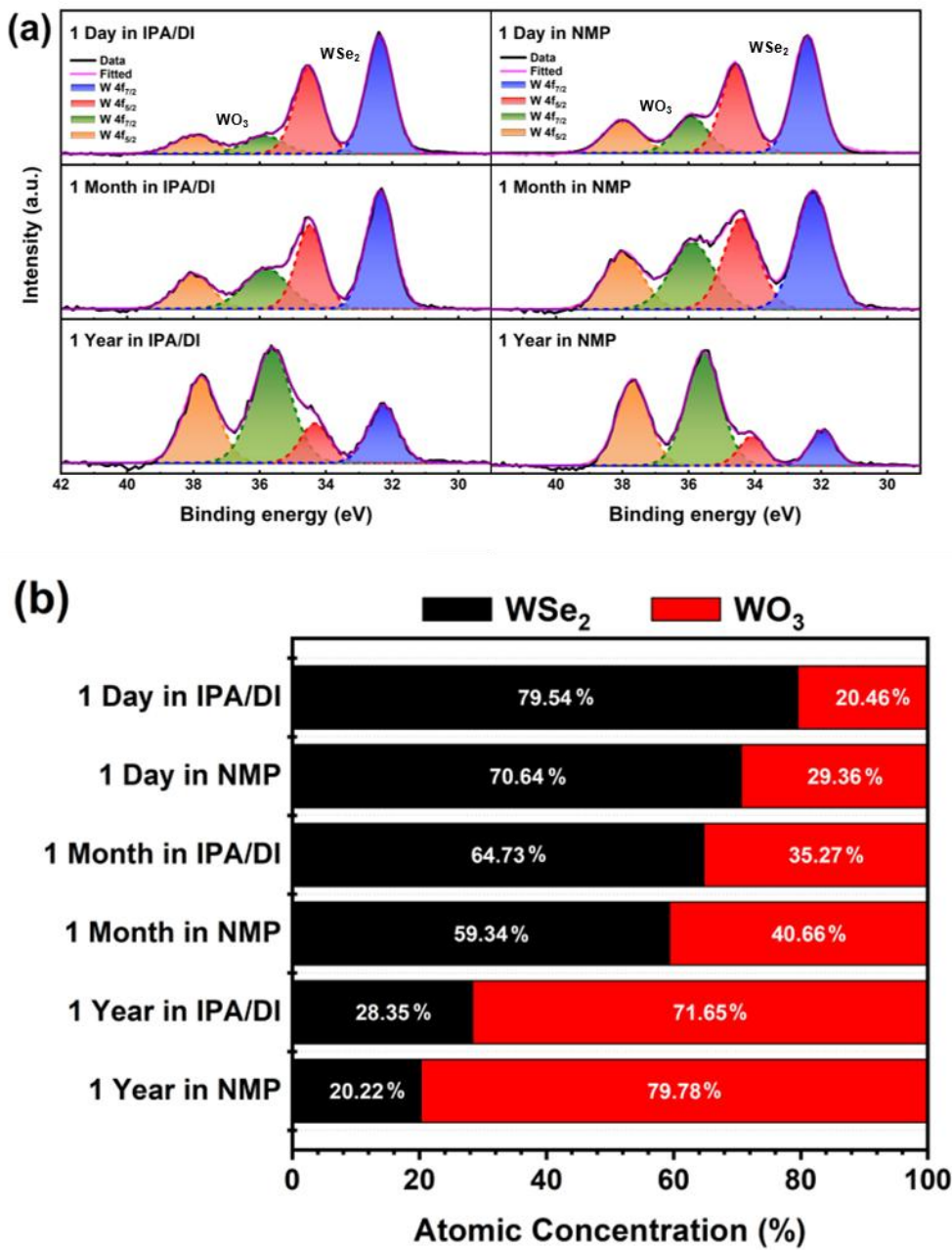

**Figure 2.** XPS analysis of the surface chemical composition of WSe$_2$ nanosheets exfoliated in IPA/DI and NMP over storage durations from 1 day to 1 year: (a) W 4f spectra and (b) atomic concentration ratios of WSe$_2$ to WO$_3$ for each sample.

While X-ray-based techniques have confirmed the time- and solvent-dependent oxidation of exfoliated $WSe_2$ nanosheets, their surface morphology and structural evolution required further investigation. To address this, TEM analysis was conducted, and the HR-TEM images of all $WSe_2$ samples are presented in Figure 3, with corresponding selected area electron diffraction (SAED) patterns shown in the insets. At a magnification of 500,000x, $WSe_2$ nanosheets exfoliated in IPA/DI (Figure 3a-c) retain clear crystallinity and exhibit the characteristic hexagonal lattice of 2H-$WSe_2$ across all storage durations. In contrast, for NMP-exfoliated samples, only the sample stored for one day (Figure 3d) maintains a visible hexagonal structure. By one month (Figure 3e), the lattice structure becomes indistinct, though the interplanar d-spacing can still be identified. After one year (Figure 3f), the lattice planes are no longer resolvable, suggesting significant structural degradation. This degradation is attributed to increased oxidation over time, especially in NMP, which promotes auto-oxidation of $WSe_2$ nanosheets. Interestingly, the relationship between oxide content and surface structure appears complex. Although the IPA/DI sample stored for one year exhibits a higher oxide content (~72%) than the NMP-exfoliated sample stored for one month (~40%), the latter shows a more significant structural loss, with the hexagonal framework no longer observable. This suggests that the NMP not only promotes oxidation but may also enhance the formation of selenium vacancies, accelerating amorphisation. These observations are further supported by SAED analysis. The insets of Figure 3a-d show distinct, bright dotted rings, characteristic of crystalline materials[19]. In the inset of Figure 3e (NMP, one month), the diffraction rings appear more diffuse and continuous. By one year (inset of Figure 3f), the rings become completely diffuse and thicker, indicating a transition to an amorphous structure[19]. This finding aligns well with the XAS results, further confirming the progressive structural disorder over time.

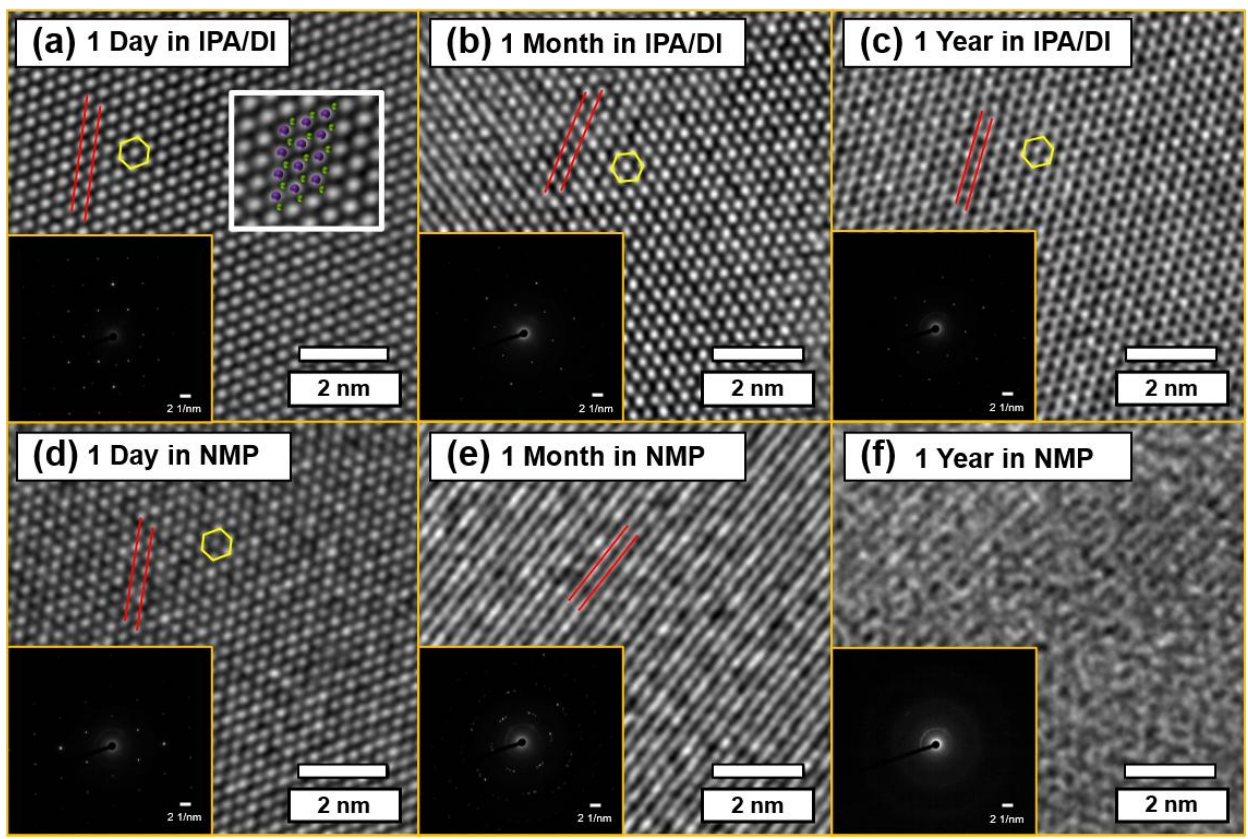

**Figure 3.** HR-TEM images of WSe$_2$ nanosheets exfoliated using IPA/DI and NMP, stored for (a) 1 day, (b) 1 month, and (c) 1 year for IPA/DI, and (d) 1 day, (e) 1 month, and (f) 1 year for NMP. The corresponding SAED patterns are shown as insets for each sample.

Based on comprehensive characterisation, WSe$_2$ nanosheets exfoliated in IPA/DI and NMP and stored for varying durations exhibit distinct local structures, surface compositions, and morphologies. This study also examines how these differences affect their HER performance to identify optimal synthesis and storage conditions. The HER polarisation curves with iR compensation, obtained via linear sweep voltammetry (LSV) in 0.5 M H$_2$SO$_4$, are shown in Figure 4a. Among all samples, the WSe$_2$ nanosheets exfoliated in IPA/DI and stored for one day (black solid line) exhibit the best HER activity, with an overpotential of -0.660 ± 0.002 V vs RHE at -10 mA cm$^{-2}$.

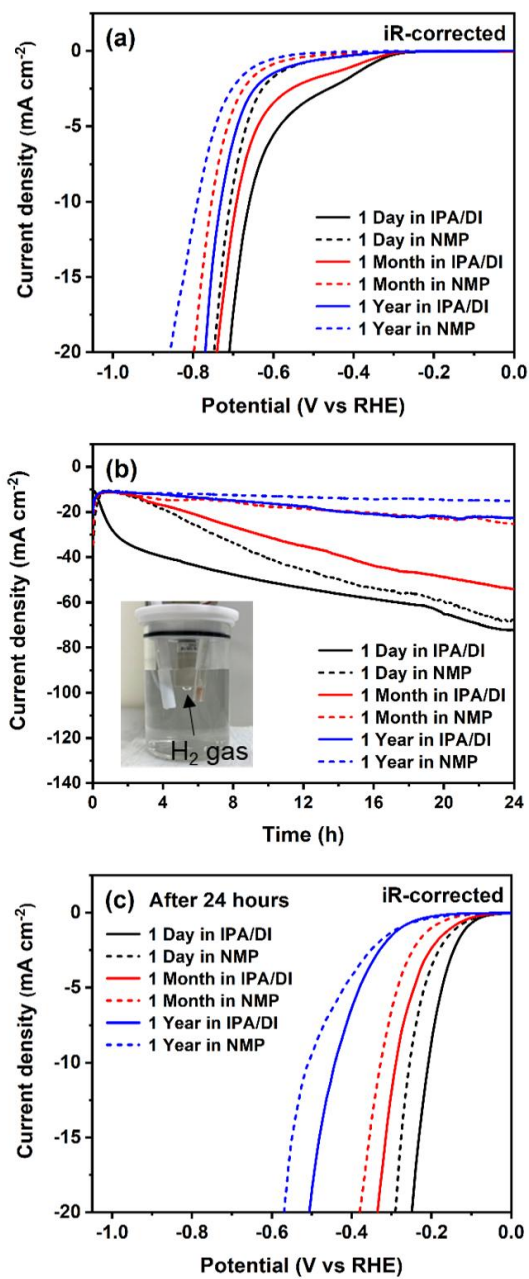

**Figure 4.** Electrocatalytic performance of as-exfoliated WSe$_2$ nanosheets in 0.5 M H2SO4: (a) *iR*-corrected polarisation curves, (b) chronoamperometry over 24 hours with the setup shown in the inset, and (c) *iR*-corrected polarisation curves post-CA for all samples.

This sample also outperforms others across various electrolytes (KOH and NaCl), as shown in the ESI. The IPA/DI-exfoliated sample stored for one month (red solid line) ranks second (-0.691 ± 0.004 V), surpassing the NMP-exfoliated sample stored for one day (black dotted line), despite having higher oxide content. This might be attributed to the influence of local atomic structure and crystallinity, as revealed by EXAFS, which shows more defined W–Se bonding in the IPA/DI-exfoliated sample. Following in order of performance are: IPA/DI-exfoliated for one year (blue solid line), NMP-exfoliated for one month (red dotted line), and finally, NMP-exfoliated for one year (blue dotted line), which exhibited the poorest HER activity with an overpotential of -0.789 ± 0.005 V vs RHE. Full EIS, iR-compensation, and overpotential data across all conditions and electrolytes are provided in Figure S5-S7 and Table S4.

Long-term stability was assessed via chronoamperometry (CA) over 24 hours in 0.5 M $H_2SO_4$ at constant potential, with initial current densities of approximately -10 mA $cm^{-2}$. As shown in Figure 4b, current density increased significantly over time, reaching around -72 mA $cm^{-2}$ for the IPA/DI-exfoliated sample stored for one day, while the NMP-exfoliated sample stored for one year showed only a modest increase to -15 mA $cm^{-2}$. Starting potentials and final current densities are summarised in Table S5. Post-CA polarisation curves (Figure 4c) reveal marked improvements in HER performance. The IPA/DI-exfoliated sample stored for one day exhibited the lowest overpotential after CA (-0.209 ± 0.010 V vs RHE), improving by ~0.45 V. In contrast, although the NMP-exfoliated sample stored for one year still showed the weakest performance, its overpotential improved to -0.508 ± 0.007 V vs RHE, a ~0.28 V gain. All samples demonstrated enhanced HER activity after CA, likely due to surface activation and increased electrochemical surface area (ECSA), which facilitate the removal of surface impurities and the exposure of additional active sites[20, 21]. Interestingly, the post-CA polarisation trend differs from the initial measurements, highlighting the role of oxide content in long-term HER performance. While oxide species can initially contribute to HER via surface oxygen or hydroxyl sites, they tend to degrade under prolonged acidic conditions, diminishing catalytic activity. In contrast, samples with lower oxide content—typically those stored for shorter durations—retain more of the WSe$_2$ phase and stable Se-edge sites, which are well-established active centres for HER. Moreover, with fewer selenium vacancies, these samples provide more favourable H adsorption (H$_{ads}$) sites, contributing to superior long-term performance[21]. Detailed overpotential values after 24 hours CA are provided in Table S5 of the ESI.

CONCLUSION

In conclusion, this study highlights the critical influence of solvent choice during LPE and storage duration on the electrocatalytic performance of the exfoliated WSe$_2$ nanosheets for the HER. A clear time- and solvent-dependent oxidation behaviour was observed, NMP-exfoliated WSe$_2$ nanosheets exhibiting greater oxidation, less-defined W–Se bonding, increasing selenium vacancies and a more amorphous structure. Additionally, longer storage times led to a significant increase in oxide content. These structural changes impact the HER activity significantly. While IPA/DI-exfoliated nanosheets initially performed best due to their well-preserved structure, after prolonged electrochemical testing, less-oxidised samples exhibited superior performance, emphasising the negative effect of excessive oxidation. All samples showed improved HER activity after 24 hours of chronoamperometry, likely due to surface activation and increased exposure of active sites. The lowest overpotential achieved was -0.206 V vs RHE at -10 mA cm$^{-2}$, comparable to platinum. Overall, these findings provide valuable insights and practical guidance for optimising the synthesis and storage conditions of WSe$_2$ nanosheets for electrocatalytic applications, particularly in HER.

## ASSOCIATED CONTENT

**Supporting Information**.

Experimental section: XAS and XPS characterisation results analysis. Electrochemical evaluation analysis: EIS analysis and HER performance evaluations in various electrolytes and long-term stability evaluation of WSe$_2$ nanosheets. The following files are available free of charge.

## AUTHOR INFORMATION

**Corresponding Author**

**Pawin Iamprasertkun** – *School of Bio-Chemical Engineering and Technology, Sirindhron International Institute of Technology, Thammasat University, Khlong Luang, Pathum Thani 12120, Thailand;* Phone: +66-2-986-9009 ext 2306; Email: pawin@siit.tu.ac.th

## ACKNOWLEDGMENT

This study was supported by Thammasat University Research Unit in Sustainable Electrochemical Intelligent (RU-SEI). The authors acknowledge the Synchrotron facilities from BL1.1W: Multiple X-ray Techniques (XAS), and BL3.2Ua: Photoemission Electron Spectroscopy (PES), Synchrotron Light Research Institute (Public Organization), Thailand.

# Supporting Information

# Table of Contents

# EXPERIMENTAL PROCEDURES

### Liquid-phase exfoliation of WSe$_2$

This study investigates the effects of solvent selection and storage duration on the properties of exfoliated WSe$_2$ nanosheets. To initiate exfoliation, 2.0 g of WSe$_2$ powder (99.8%, Alfa Aesar) was dispersed in 200 mL of solvent. For the NMP system, the powder was added to 200 mL of N-methyl-2-pyrrolidone (98%, Loba). For the mixed solvent system, WSe$_2$ powder was added to a 1:1 (v/v) mixture of 100 mL isopropyl alcohol (IPA) and 100 mL deionised water (Milli Q). The resulting suspensions were sonicated in an ultrasonic bath at a power of 271.3 W for 12 hours at 25°C. Following the exfoliation, the dispersions were centrifuged at 5000 rpm for 20 minutes. The exfoliated WSe$_2$ nanosheets were collected from the supernatant, while the unexfoliated residues remained at the bottom of the tubes. The collected nanosheets were then stored in a cabinet for various durations—1 day, 1 month, and 1 year—to study the effect of storage time. For solid-state characterisation and nanosheets concentration, the exfoliated WSe$_2$ dispersions were filtered onto PVDF membranes. Omnipore membrane (polyvinylidene fluoride, PVDF, hydrophilic surface, 0.1 μm pore size) was purchased from Merck Millipore Limited. Before filtration, the membranes were pre-weighed using a five-digit precision balance (Mettler Toledo). The dispersions were filtered using a syringe pump at a rate of 20 mL h$^{-1}$. To prevent further oxidation due to exposure to atmospheric oxygen and moisture, the resulting WSe$_2$ thin films were stored in a desiccator until further use.

### Material characterisations

The thin WSe$_2$ films deposited on PVDF membranes were utilised for various characterisation techniques requiring solid-state samples. X-ray diffraction (XRD) analysis was performed using a Bruker D8 Advance diffractometer (Bruker AXS GmbH, Germany) equipped with a Cu-Kα radiation source (λ = 1.5406 Å). Scans were conducted over a 2θ range of 10–80°, with a step size of 0.02° and a counting time of 0.2 seconds per step. X-ray absorption near edge structure (XANES) measurements were carried out at Beamline 1 of the Synchrotron Light Research Institute (SLRI), Nakhon Ratchasima, Thailand, to probe the local chemical environment of tungsten. For solid-state measurements, dispersions containing comparable quantities of WSe$_2$ nanosheets were filtered onto PVDF membranes and measured in transmission mode. For liquid-state analysis, 0.3 mL of exfoliated WSe$_2$ dispersion was loaded into a polyamide 6 (PA6) liquid cell holder sealed with X-ray transparent Kapton tape. The W L-edge spectra were acquired using photon energy scan points of –150, –20, 60, 100, 200, and 12,000 eV, with corresponding step sizes of 3, 0.3, 1, 3, and 500 eV and dwell times of 2, 2, 3, 5, and 6 seconds, respectively. The total acquisition time per scan was approximately 40 minutes. Due to the low concentration of absorbing atoms in the dispersions, fluorescence detection mode was employed to enhance signal sensitivity. X-ray photoelectron spectroscopy (XPS) was also conducted at SLRI using a SCIENTA R4000 hemispherical analyser at the BL3.2Ua beamline. Surface chemical composition was analysed using photon energies in the ranges of 40–160 eV and 240–1040 eV. For transmission electron microscopy (TEM), the WSe$_2$ dispersions were diluted with isopropanol, drop-cast onto carbon-coated copper grids, and dried in a vacuum oven. TEM imaging was performed using a JEOL JEM-2010F microscope operated at an accelerating voltage of 200 kV.

### Electrochemical measurements

For electrode preparation, 100 µL of each WSe$_2$ dispersion (~1.0 mg mL$^{-1}$) was mixed thoroughly with 4 µL of Nafion solution (5 wt% in lower aliphatic alcohols and water, Sigma-Aldrich). The resulting mixture was drop-cast onto a polished glassy carbon electrode (surface area = 0.0707 cm$^2$) and dried in an oven at 55°C for 6 hours. These prepared electrodes were used as the working electrodes for standard electrocatalytic measurements. A three-electrode setup was employed, comprising the WSe$_2$-coated glassy carbon electrode as the working electrode (WE), a double-junction Ag/AgCl reference electrode (RE), and a polycrystalline platinum wire as the counter electrode (CE), all sourced from PalmSens.

To investigate the electrocatalytic performance under different pH conditions, three types of electrolytes were used. These included an acidic electrolyte of 0.5 M H$_2$SO$_4$ with a pH of 0.703, a basic electrolyte of 1.0 M KOH with a pH of 13.43, and a set of neutral electrolytes prepared with varying concentrations of NaCl to simulate seawater conditions. The NaCl concentrations used were 1.0 M, 0.1 M, 0.01 M, and 0.001 M, corresponding to pH values ranging from 6.77 to 5.76. All electrolyte solutions were prepared by dissolving the corresponding materials in Milli-Q deionised water. Prior to all electrochemical tests, the electrolyte solutions were purged with nitrogen gas for 30 minutes to eliminate dissolved oxygen.

For long-term stability testing, 0.5 M H$_2$SO$_4$ was chosen as the electrolyte due to the superior HER activity observed in this medium. A rotating disk electrode (RDE) system was employed, using a glassy carbon disk electrode with the same working area of 0.0707 cm² (ALS, Japan) as the working electrode. The WSe$_2$ dispersion mixed with Nafion was drop-cast onto the RDE using the same procedure as for the standard glassy carbon electrode. A carbon rod was used as the counter electrode in place of platinum to prevent potential Pt deposition on the WSe$_2$ surface, which could artificially enhance HER activity. The Ag/AgCl electrode continued to serve as the reference electrode, supported by prior reports indicating that silver leakage from Ag/AgCl does not contribute to improved HER performance[1].

Linear sweep voltammetry (LSV) was used to assess the HER kinetics, with a scan rate set at 5 mV s$^{-1}$. All measured potentials were calibrated to the reversible hydrogen electrode (RHE) using the equation: $E(RHE) = E(Ag/AgCl) + 0.197 + 0.059 \times pH$.[2] To ensure accurate performance analysis, all polarisation curves were corrected for iR drop using 100% solution resistance compensation[3]. The resistance values used for iR correction were obtained through electrochemical impedance spectroscopy (EIS) conducted over a frequency range of 0.01 to $10^5$ Hz.

# CHARACTERISATION RESULTS

### X-ray absorption spectroscopy (XAS) analysis

The normalised W $L_3$-edge XANES spectra of all as-exfoliated $WSe_2$ samples are presented in Fig. S1, with corresponding white line peak positions summarised in Table S1. A noticeable shift in the white line peaks toward higher energies, along with a reduction in peak intensity, was observed for the nanosheets exfoliated in NMP. Additionally, an increase in storage duration led to a further upward shift in peak positions. As shown in Table S1, the white line peaks of $WSe_2$ nanosheets exfoliated in IPA/DI and stored for one day and one month, as well as those exfoliated in NMP and stored for one day, appear at approximately 10,210 eV—consistent with the $W^{4+}$ oxidation state. In contrast, the samples exfoliated in IPA/DI and stored for one year, and those exfoliated in NMP and stored for one month and one year, exhibit peaks shifted to around 10,212 eV, indicative of partial oxidation to $W^{6+}$. These values align with the previously published results[4, 5]. This energy shift reflects a clear time- and solvent-dependent oxidation behaviour, where nanosheets exfoliated in NMP undergo more significant oxidation, which also increases progressively with storage time. The observed reduction in white line intensity suggests possible structural disorder or changes in the local coordination environment associated with higher degrees of oxidation. To further explore these structural variations, an EXAFS analysis was conducted and is discussed in the following section.

**Table S1.** White line peak positions for all as-exfoliated $WSe_2$ samples.

|  | White line peak position | |
| --- | --- | --- |
| **Storage duration** | **Exfoliated in IPA/DI** | **Exfoliated in NMP** |
| 1 day | 10210.2 eV | 10210.3 eV |
| 1 month | 10210.2 eV | 10211.9 eV |
| 1 year | 10212.2 eV | 10212.3 eV |

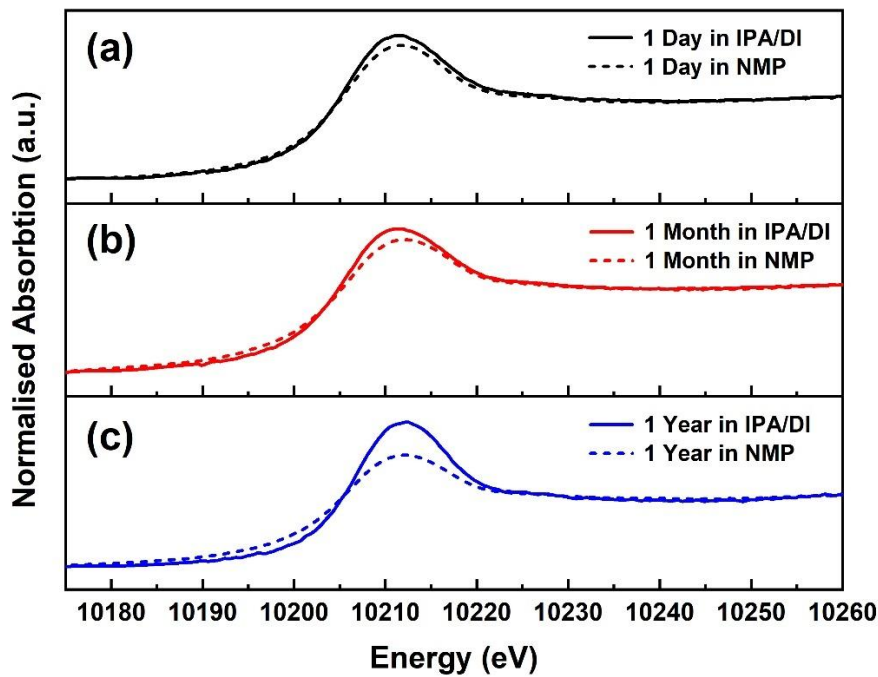

**FigS. 1** Normalised W L$_3$-edge XANES spectra of WSe$_2$ nanosheets exfoliated in IPA/DI (solid lines) and NMP (dotted lines), after storage durations of (a) 1 day, (b) 1 month, and (c) 1 year.

The Fourier transform (FT) curves derived from the EXAFS spectra of all WSe$_2$ samples in liquid dispersion form are presented in Fig. S2. For the nanosheets exfoliated in IPA/DI and stored for one day (Fig. S2a) and one month (Fig. S2b), the FT spectra exhibit a dominant W–Se peak located at approximately 2.1 Å, while the first-shell W–O peak appears at around 1.46 Å. Both peaks remain well-defined and consistent in position, indicating minimal structural change over this period. However, after one year of storage (Fig. S2c), a noticeable shift of both peaks to higher radial distances is observed, accompanied by a significant increase in the intensity of W–O peak, which becomes predominant. Despite this, the W–Se peak remains distinguishable. This trend suggests progressive oxidation of the WSe$_2$ nanosheets over time, leading to the increased formation of W–O bonds. In contrast, nanosheets exfoliated in NMP show a dominant W–O peak even after just one day of storage (Fig. S2d), indicating more immediate and extensive oxidation in this solvent. Although the W–Se peak is still visible at this stage, it becomes increasingly diminished over time. By one month (Fig. S2e) and one year (Fig. S2f), the W–O peak overwhelmingly dominates, and the W–Se peak becomes barely discernible. In addition to changes in peak intensities, a consistent shift of peak positions toward higher radial distances is evident, as detailed in Table S2. This shift, particularly pronounced in the NMP-exfoliated samples and increasing with storage duration, reflects elongation of the W–Se bond. The accompanying decrease in W–Se peak intensity suggests the introduction of selenium vacancies and structural distortion[5], contributing to the amorphisation of the nanosheets. These observations confirm the time- and solvent-dependent oxidation behaviour of WSe₂ nanosheets, with NMP promoting more rapid and extensive degradation of the local atomic structure compared to the IPA/DI system.

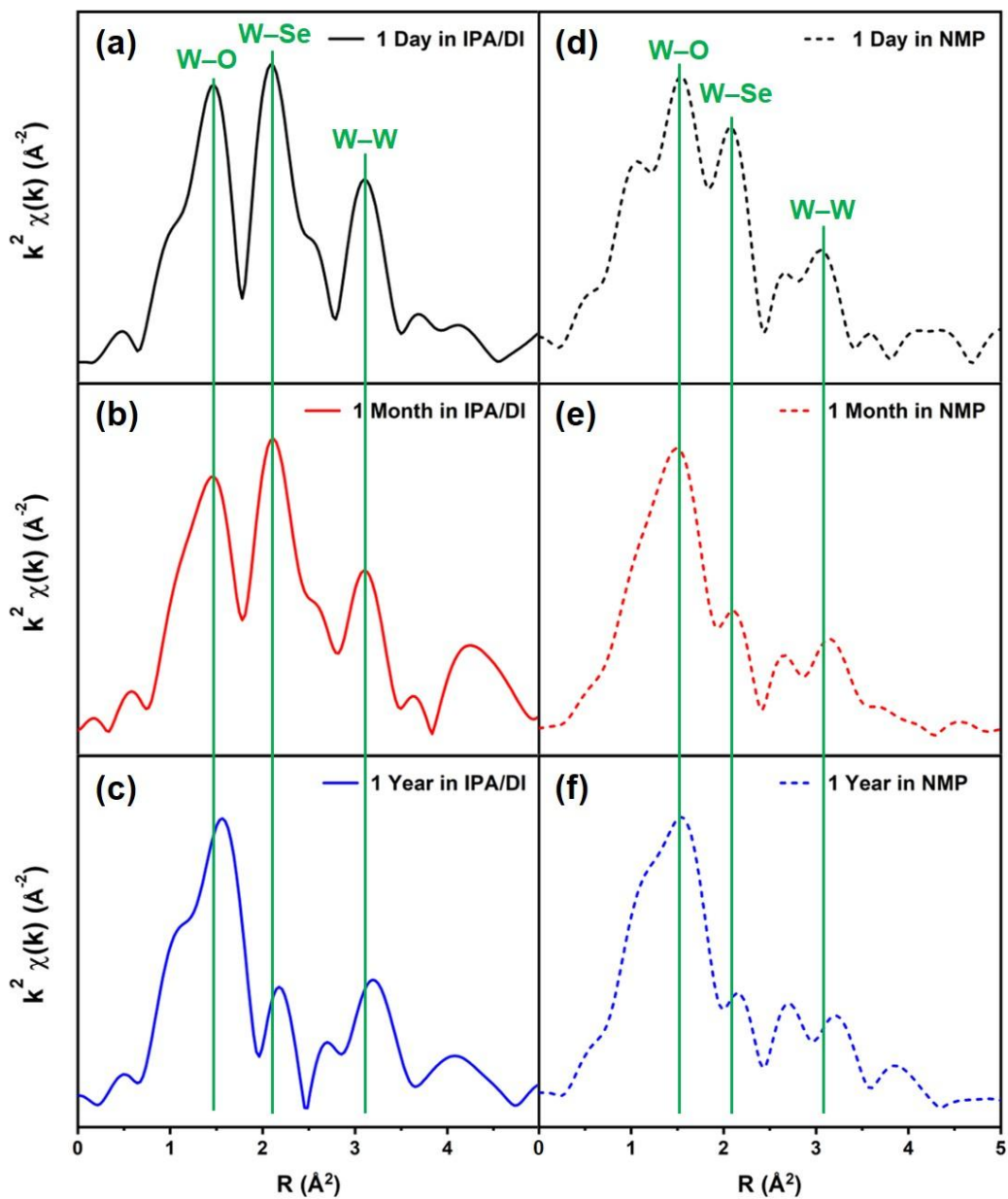

**Fig. S2** Fourier-transformed EXAFS spectra showing the radial distribution functions of WSe$_2$ nanosheets exfoliated in IPA/DI and stored for (a) 1 day, (b) 1 month, and (c) 1 year, and in NMP for (d) 1 day, (e) 1 month, and (f) 1 year.

**Table S2.** Peak positions derived from the EXAFS spectra of all as-exfoliated WSe$_2$ samples.

|  | Radial distances | | | |
|---|---|---|---|---|
|  | **Exfoliated in IPA/DI** | | **Exfoliated in NMP** | |
| **Storage duration** | **W–O** | **W–Se** | **W–O** | **W–Se** |
| 1 day | 1.46 Å | 2.08 Å | 1.53 Å | 2.09 Å |
| 1 month | 1.47 Å | 2.11 Å | 1.52 Å | 2.11 Å |
| 1 year | 1.56 Å | 2.18 Å | 1.55 Å | 2.17 Å |

XAS measurements were also conducted on solid-state WSe$_2$ nanosheets, prepared as thin films by filtering the dispersions onto the PVDF membranes. This was done to enable the comparison between the nanosheets in the solid and dispersion states. In the XANES spectra (Fig. S3a-c), shifts in the white line peak positions and reductions in intensity were observed, particularly for samples exfoliated in NMP across all storage durations. However, these changes were less pronounced than those seen in the dispersion-state measurements. Differences in local atomic structure are more clearly revealed in the EXAFS spectra, shown in Fig. S3d-f. Unlike the dispersion-state spectra, where the first-shell peak corresponds to W–O bond, the solid-state samples—especially those exfoliated in IPA/DI—exhibit a dominant first-shell peak at approximately 2.07 Å, corresponding to the W–Se bond. In samples exfoliated in NMP, this peak shifts slightly to a longer radial distance of around 2.12 Å. Across all samples, the W–Se bond peaks are generally more prominent than the W–O peaks, with the exception of the NMP-exfoliated samples stored for one month (Fig. S3e, dotted red line), where both peaks have comparable intensity, and for one year (Fig. S3f, dotted blue line), where the W–O peak dominates. The distinct W–Se peaks, alongside the broad and faint W–O signals in the solid-state spectra, can be attributed to the re-stacking of nanosheets during the drying process on the PVDF membrane. In contrast, in the dispersion state, the nanosheets remained well-separated and surrounded by the solvent molecules, allowing oxygen atomic to interact more readily with the W atoms and thus producing more pronounced W–O features[6]. Despite these differences, the time- and solvent-dependent oxidation behaviour of the WSe$_2$ nanosheets remains evident in the solid-state measurements. W–O bond features are more pronounced in the NMP-exfoliated samples, and the W–Se peaks gradually weaken with increased storage time, indicating progressive oxidation and structural degradation.

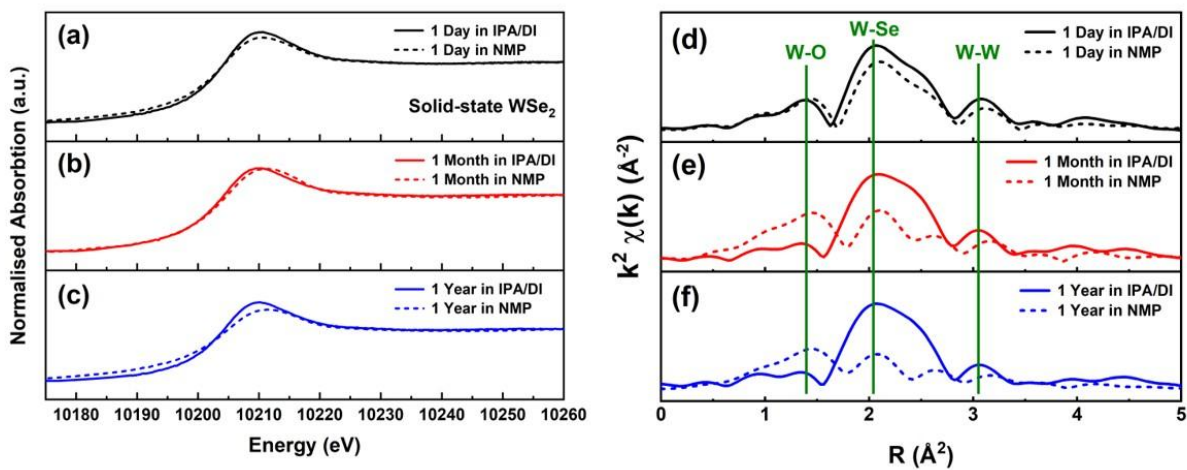

**Fig. S3** Solid-state XAS measurements of WSe$_2$ nanosheets. (a–c) XANES spectra and (d–f) EXAFS spectra of samples exfoliated in IPA/DI and NMP and stored for 1 day, 1 month, and 1 year, respectively.

### X-ray photoelectron spectroscopy (XPS) analysis

To analyse the surface chemical composition of the as-exfoliated WSe$_2$ nanosheets, XPS measurements were conducted. The W 4f spectra of all samples are presented in Fig. S4a. A clear increase in the intensity of the W $4f_{7/2}$ and W $4f_{5/2}$ peaks corresponding to WO$_3$ (W$^{6+}$) is observed with increasing storage time, particularly in samples exfoliated in NMP compared to those in IPA/DI. Notably, a shift toward lower binding energies is also evident in samples with higher oxide content. The binding energy values of W 4f spectra for each sample are summarised in Table S3. A similar trend is observed in the Se 3d region, as shown in Fig. S4b. All samples exhibit Se 3d doublet peaks between 53 and 57 eV, corresponding to Se $3d_{5/2}$ at approximately 54.7 eV and Se $3d_{3/2}$ at around 55.4 eV. Detailed binding energy values are also listed in Table S3. As discussed in the main text, the decreasing intensity of WSe$_2$-related peaks and increasing intensity of WO$_3$-related peaks reflect a time- and solvent-dependent oxidation process. Samples exfoliated in NMP and stored for longer periods exhibit the highest degree of oxidation. In addition to peak intensity changes, the shift in binding energy provides further insight into the oxidation state. Previous studies report binding energies of 32.3 and 34.4 eV for W $4f_{7/2}$ and W $4f_{5/2}$ peaks in pristine 2H-WSe$_2$, respectively. Upon oxidation, these peaks shift to lower binding energies, while new peaks emerge at ~35.6 eV and ~37.7 eV, corresponding to the W 4f doublet of WO$_3$. This trend aligns with literature findings, indicating that as the oxidation level increases (WO$_X$, where $x \leq 3$), the W 4f and Se 3d binding energies progressively shift downward[7, 8]. In summary, the XPS results confirm that WSe$_2$ nanosheets exfoliated in NMP undergo more extensive oxidation, and the degree of oxidation increases with longer storage time, as evidenced by the emergence and intensification of WO$_3$ peaks and the corresponding binding energy shifts.

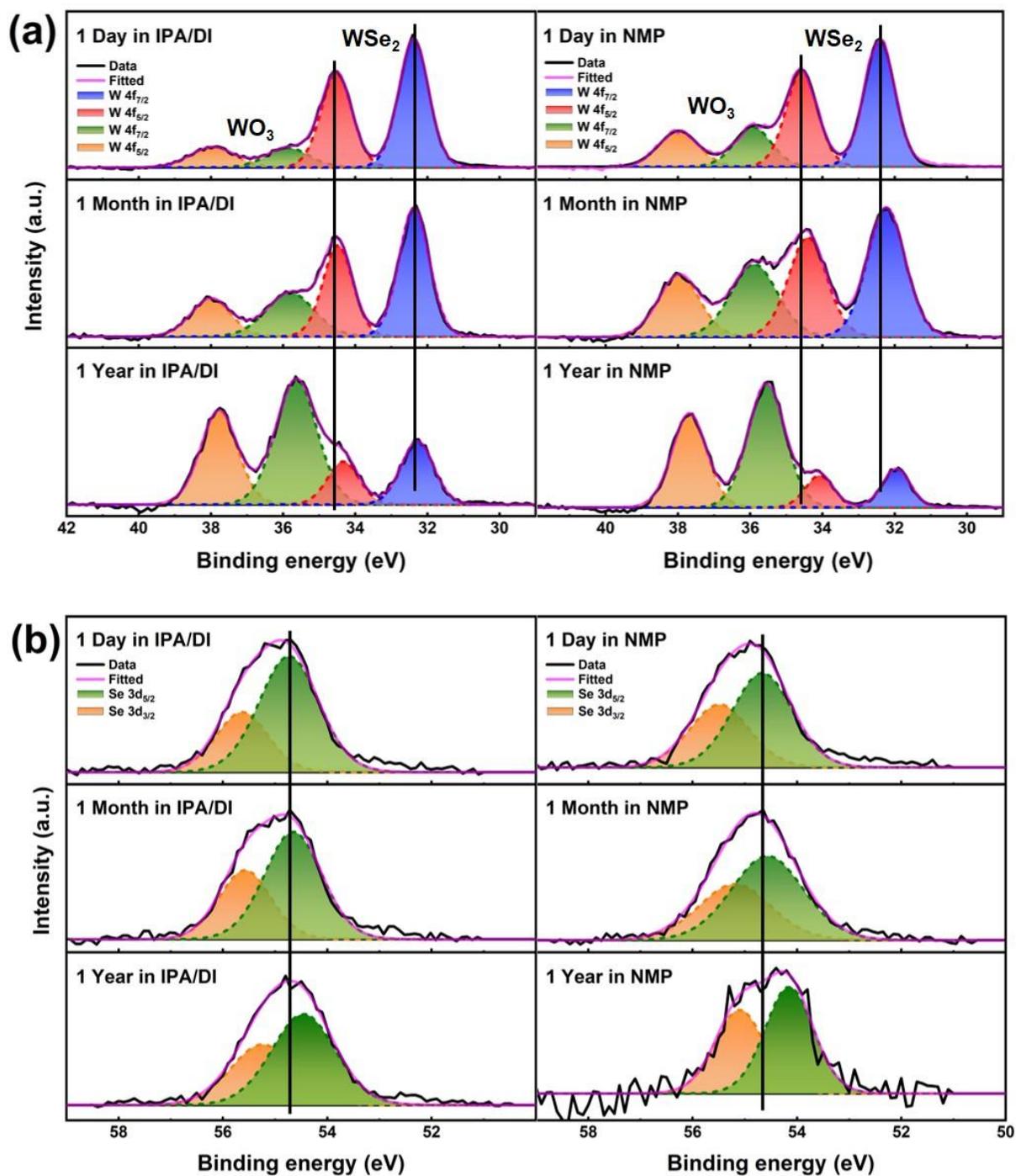

**Fig. 4** X-ray photoelectron spectroscopy (XPS) analysis of the as-exfoliated WSe$_2$ nanosheets: (a) W 4f spectra and (b) Se 3d spectra of all samples.

**Table S3.** Binding energy values of the W 4f and Se 3d spectra for all as-exfoliated WSe$_2$ samples.

|  |  | Binding energies (eV) | | | | | |
|---|---|---|---|---|---|---|---|
| **Storage duration** | **Solvent used** | WSe$_2$ | | WO$_3$ | | Se | |
|  |  | W 4f$_{7/2}$ | W 4f$_{5/2}$ | W 4f$_{7/2}$ | W 4f$_{5/2}$ | Se 3d$_{5/2}$ | Se 3d$_{3/2}$ |
| 1 day | IPA/DI | 32.3 | 34.4 | 35.7 | 37.9 | 54.7 | 55.5 |
|  | NMP | 32.3 | 34.4 | 35.7 | 37.9 | 54.6 | 55.4 |
| 1 month | IPA/DI | 32.2 | 34.4 | 35.7 | 37.8 | 54.6 | 55.5 |
|  | NMP | 32.2 | 34.3 | 35.7 | 37.8 | 54.5 | 55.1 |
| 1 year | IPA/DI | 32.2 | 34.2 | 35.6 | 37.7 | 54.4 | 55.1 |
|  | NMP | 31.9 | 34.0 | 35.4 | 37.5 | 54.1 | 55.0 |

## ELECTROCATALYTIC PROPERTIES ANALYSIS

### Electrochemical impedance spectroscopy (EIS) analysis

To correct the voltage loss arising from the electrolyte resistance between the working and reference electrodes, iR compensation was applied[9]. EIS was conducted to determine the solution resistance ($R_s$), which was obtained by fitting the high-frequency region of the Nyquist plot using an equivalent circuit model. In this study, a Randles circuit was employed for the fitting. Fig. S5a displays the Nyquist plot of the WSe$_2$ nanosheets exfoliated in IPA/DI and stored for one day, shown here as a representative sample. The plot includes both the experimental data and the fitted curve, with the equivalent circuit diagram in the inset. The Nyquist plot exhibits a nearly vertical line, characteristic of capacitive behaviour[10], and notably lacks a semicircle. This absence of the semicircular feature indicates minimal charge storage distribution and strong Ohmic contact between the WSe$_2$ nanosheets and the substrate[11, 12]. Similar Nyquist plot profiles were observed for all as-exfoliated WSe$_2$ samples across all electrolytes tested. This capacitive response is consistent with previously reported Nyquist plots for WSe$_2$ and WO$_3$ nanosheets[10, 13]. Given the lack of a visible semicircle, implying negligible charge transfer resistance ($R_{ct}$) and fast electron transfer kinetics, 100% iR compensation was deemed appropriate. Under these conditions, the uncompensated resistance ($R_u$) can reasonably be assumed to represent the solution resistance ($R_s$) alone[3]. The $R_s$ values for each sample were extracted from the high-frequency intercept of their respective Nyquist plots. Fig. S5b-g present the $R_s$ values across various electrolytes. No specific trend in the $R_s$ was observed among different WSe$_2$ samples within the same electrolytes. However, a clear trend was evident across electrolytes: the lowest $R_s$ values, approximately 10 Ω, were observed in 0.5 M H$_2$SO$_4$ (Fig. S5b) and 1.0 M KOH (Fig. S5c). In contrast, R$_s$ values in NaCl solutions increased significantly with decreasing concentration, with approximate values of 25 Ω, 150 Ω, 1000 Ω, and 5000–10,000 Ω for 1 M, 0.1 M, 0.01 M, and 0.001 M NaCl, respectively. This sharp increase in resistance at lower NaCl concentrations is attributed to the high resistivity of deionised water.

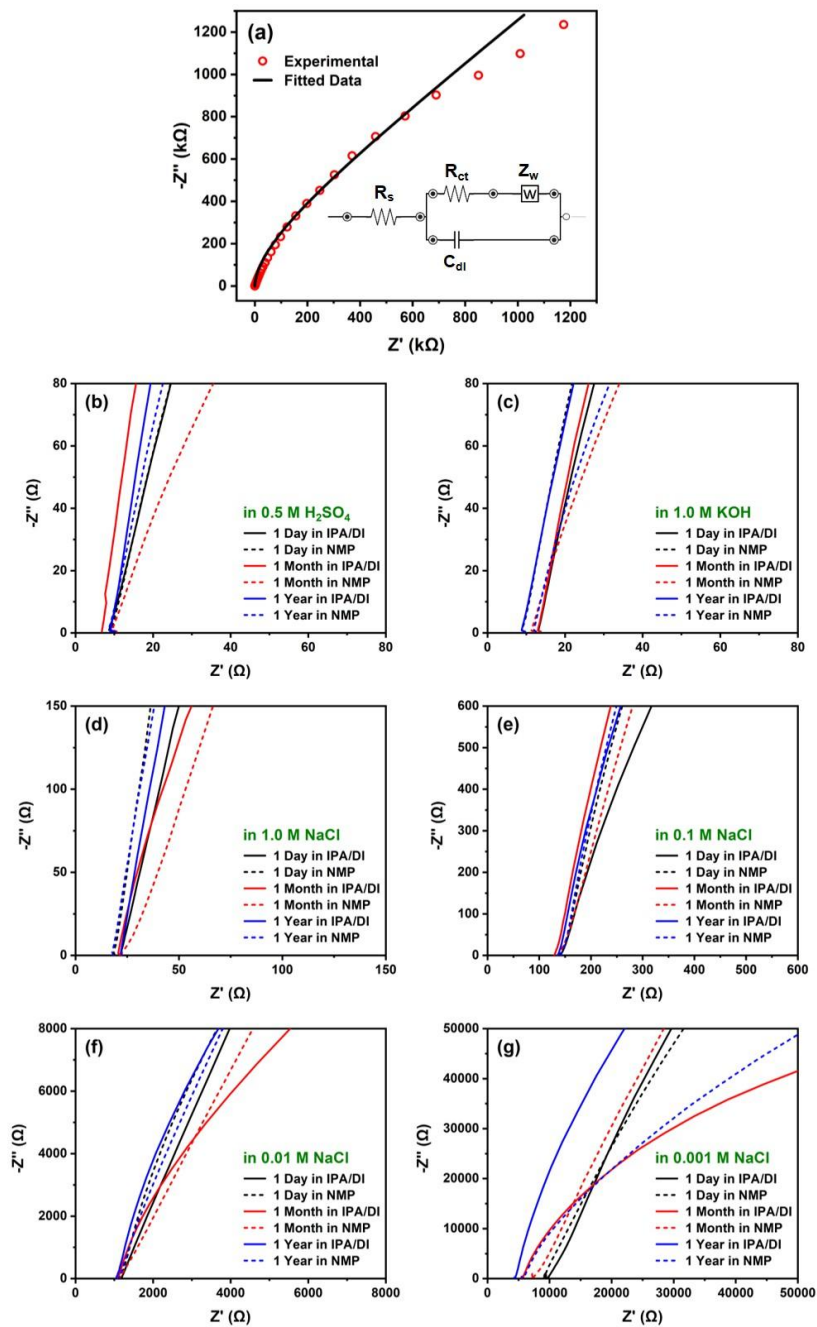

**Fig. S5** (a) Representative Nyquist plot of WSe$_2$ nanosheets exfoliated in IPA/DI and stored for 1 day, showing both experimental data and fitted curve; the inset displays the equivalent circuit used for fitting. Nyquist plots of all as-exfoliated WSe₂ samples in (b) 0.5 M H₂SO₄, (c) 1 M KOH, (d) 1 M NaCl, (e) 0.1 M NaCl, (f) 0.01 M NaCl, and (g) 0.001 M NaCl.

**Evaluation of HER catalytic activity in various electrolytes**

To assess the HER catalytic performance of the as-exfoliated WSe₂ nanosheets under different conditions, LSV was conducted in various electrolytes, including 0.5 M H₂SO₄, 1.0 M KOH, and NaCl solutions of varying concentrations. Fig. S6 presents the polarisation curves of all WSe₂ samples in these electrolytes, recorded without iR compensation to evaluate their intrinsic HER activity. Across all electrolytes, the WSe₂ samples followed a consistent performance trend based on their structural integrity and degree of oxidation. Samples with more well-defined W–Se bonds, fewer selenium vacancies, and better crystallinity exhibited superior HER activity. At the benchmark current density of -10 mA cm$^{-2}$, the best performance was observed in the sample exfoliated with IPA/DI and stored for 1 day, followed closely by the sample stored for 1 month under the same conditions. The next highest activity was recorded for the NMP-exfoliated sample stored for 1 day. These were followed in descending order by the IPA/DI-exfoliated sample stored for 1 year, then the NMP-exfoliated sample stored for 1 month, and lastly, the NMP-exfoliated sample stored for 1 year, which exhibited the lowest HER activity. This trend remained consistent across all electrolyte systems. The highest electrocatalytic activity was observed in 0.5 M H₂SO₄ (Fig. S6a), with overpotential values at -10 mA cm$^{-2}$ summarised in Table S4. While the HER activity in 1.0 M KOH (Fig. S6b) was slightly lower, it remained acceptable. The best-performing sample in KOH, WSe₂ exfoliated in IPA/DI and stored for 1 day, displayed an overpotential of -0.805 ± 0.005 V vs RHE (uncorrected). In NaCl electrolytes, HER performance progressively declined with decreasing salt concentration. At the lowest concentration of 0.001 M NaCl (Fig. S6f), none of the WSe₂ samples reached -10 mA cm$^{-2}$, even at potentials as negative as -2.5 V vs RHE. This degradation in performance is attributed to the high resistivity of the diluted NaCl solutions, which limits the current response despite increasing applied voltage. These finding suggest that highly diluted NaCl is unsuitable as an electrolyte for HER applications involving WSe₂ nanosheets.

As previously discussed, 100% iR compensation was applied to all systems. In both 0.5 M H₂SO₄ (Fig. S7a) and 1.0 M KOH (Fig. S7b), the compensation has minimal impact, with only slight reductions in overpotential. Corrected values for 0.5 M H₂SO₄ are presented in Table S4 to compare pre- and post-compensation performance. For NaCl electrolytes at moderate concentrations (1.0 M and 0.1 M; Fig. S7c and S7d), 100% iR correction led to noticeable shifts in polarisation curves toward more favourable potentials. However, in the highly diluted NaCl systems (0.01 M and 0.001 M; Fig. S7e and S7f), full compensation caused significant distortion of the polarisation curves, resulting in a "bend-back" effect and non-physical data. This artifact arises from excessive voltage correction, where the high uncompensated resistance ($R_u$ > 1000 Ω) leads to iR correction values exceeding the practical potential range of the system[3, 9]. For instance, at 10 mA and 1000 Ω, a 10 V correction would be applied. This value is well beyond the electrochemical window of the practical LSV, producing unrealistic current-voltage behaviour. In such cases, full iR compensation is not viable, and partial correction or no correction should be used instead[3]. Nevertheless, given the inherently poor HER performance in these highly resistive NaCl electrolytes, their use is not recommended for further WSe$_2$-based electrocatalytic studies.

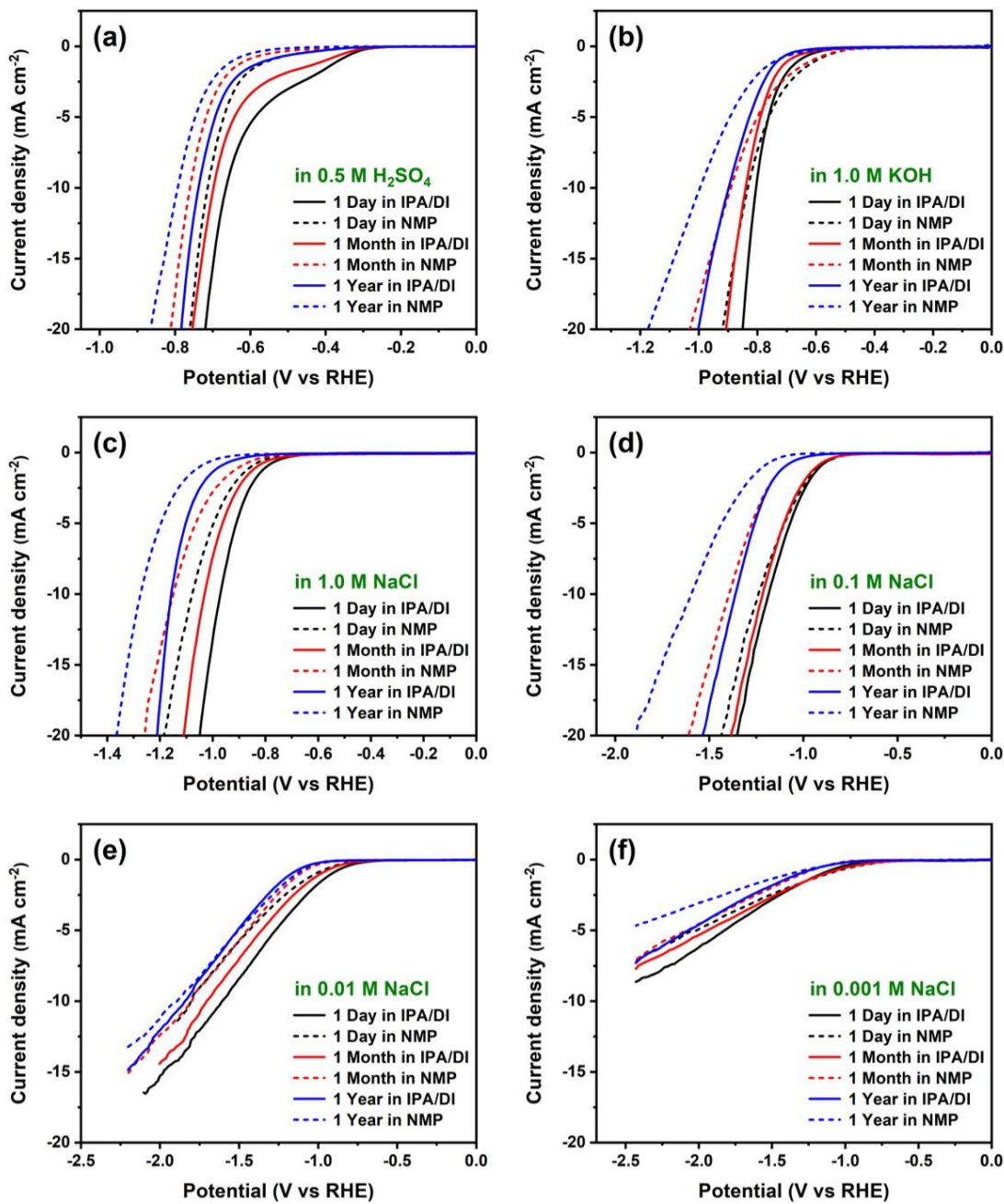

**Fig. S6** Polarisation curves without iR compensation for all as-exfoliated WSe$_2$ samples in (a) 0.5 M H$_2$SO$_4$, (b) 1.0 M KOH, (c) 1.0 M NaCl, (d) 0.1 M NaCl, (e) 0.01 M NaCl, and (f) 0.001 M NaCl.

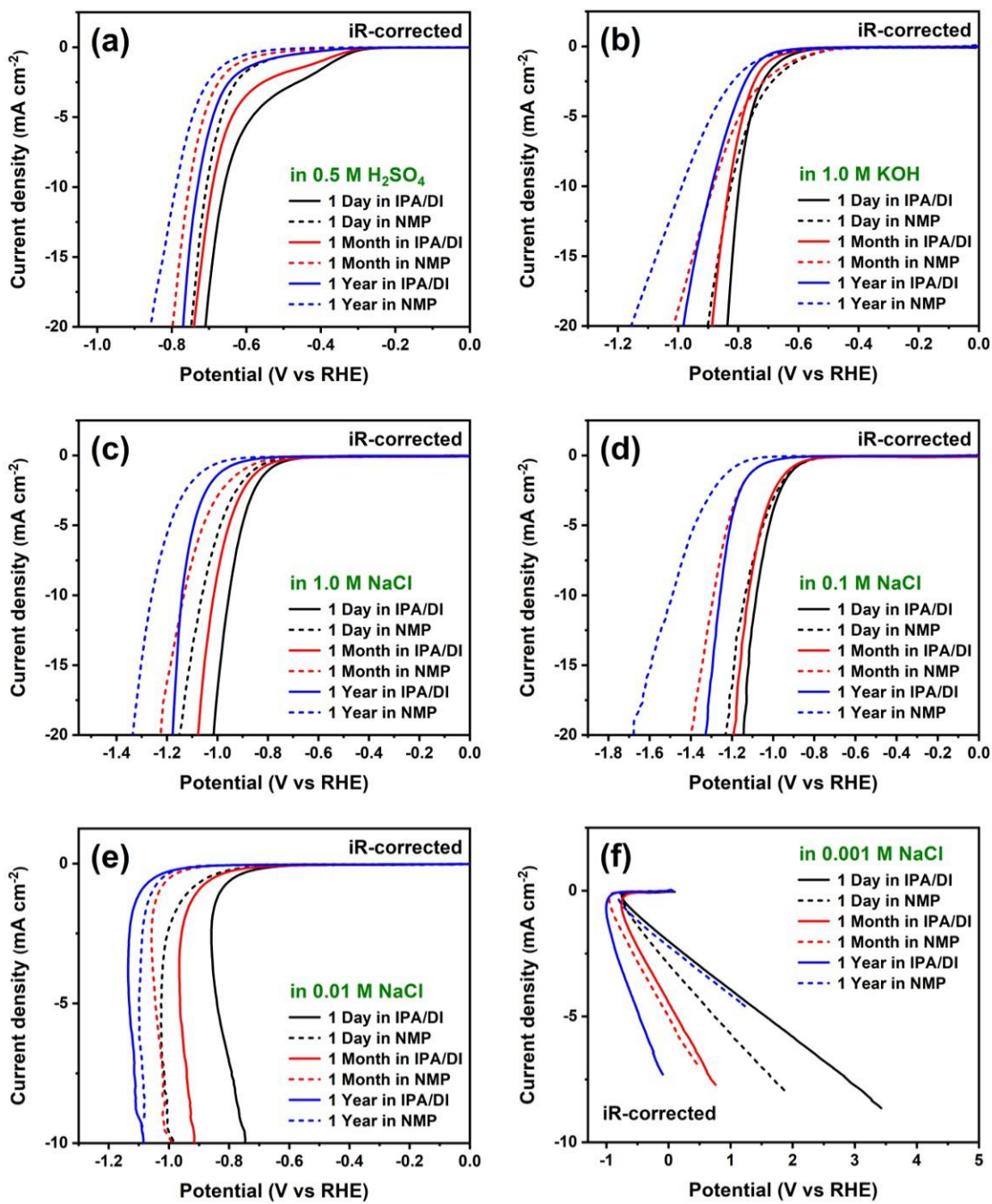

**Fig. S7** Polarisation curves with 100% iR compensation for all as-exfoliated WSe$_2$ samples in (a) 0.5 M H$_2$SO$_4$, (b) 1.0 M KOH, (c) 1.0 M NaCl, (d) 0.1 M NaCl, (e) 0.01 M NaCl, and (f) 0.001 M NaCl.

**Table S4.** Overpotential values at -10 mA cm$^{-2}$ with and without iR compensation in 0.5 M H$_2$SO$_4$ for all as-exfoliated WSe$_2$ samples.

| Overpotential values (V vs RHE) in 0.5 M H$_2$SO$_4$ at -10 mA cm$^{-2}$ | | | |
|---|---|---|---|
| **Solvent used** | **Storage duration** | **No iR compensation** | **100% iR compensation** |
| IPA/DI | 1 day | -0.665 ± 0.001 | -0.660 ± 0.002 |
| IPA/DI | 1 month | -0.698 ± 0.004 | -0.691 ± 0.004 |
| IPA/DI | 1 year | -0.738 ± 0.010 | -0.731 ± 0.009 |
| NMP | 1 day | -0.714 ± 0.003 | -0.707 ± 0.003 |
| NMP | 1 month | -0.760 ± 0.005 | -0.754 ± 0.004 |
| NMP | 1 year | -0.793 ± 0.005 | -0.789 ± 0.005 |

### Long-term stability evaluation of WSe$_2$ nanosheets.

The long-term stability of the as-prepared electrocatalysts is a critical factor for practical HER applications. Chronoamperometry (CA) measurements were conducted over a 24-hour period at a constant potential, targeting a current density of approximately -10 mA cm$^{-2}$ for each WSe$_2$ sample. These stability tests were performed exclusively in 0.5 M H$_2$SO$_4$, as this electrolyte demonstrated the highest HER catalytic activity among those tested. As noted previously, all WSe$_2$ samples exhibited an increase in current density over the 24-hour CA duration, indicating progressive surface activation. Table S5 presents the initial overpotential values and the corresponding final current densities for each sample after a 24-hour test. To assess the electrocatalytic performance following prolonged operation, LSV was also conducted post-CA. The iR-corrected overpotential values at -10 mA cm$^{-2}$ obtained from these LSV measurements are also summarised in Table S5. Among all samples, the WSe$_2$ nanosheets exfoliated in IPA/DI and stored for one day exhibited the greatest improvement in HER performance, with the reduction in overpotential of approximately 0.45 V (nearly threefold), indicating enhanced catalytic activity. Although the WSe$_2$ nanosheets exfoliated in NMP and stored for one year continued to show the poorest overall HER performance, they also showed a substantial improvement, with a decrease in overpotential of about 0.28 V. This resulted in a post-CA overpotential of -0.508 V vs RHE, which was still higher than the initial overpotential (-0.660 V) of the IPA/DI-exfoliated sample stored for one day. These improvements across all samples can be attributed to increased electrochemical active surface area and the removal of surface impurities during prolonged operation. This enhanced exposure of active sites facilitates more efficient electron transfer, thereby contributing to the observed enhancement in HER catalytic activity.

Table S5. Overpotential values at the start and iR-corrected values after 24-hour CA, along with final current densities, for all as-exfoliated $WSe_2$ samples in 0.5 M $H_2SO_4$.

| Storage duration | Solvent used | Chronoamperometry | | Linear sweep voltammetry | |
|---|---|---|---|---|---|
| | | Initial overpotential (V vs RHE) | Final current densities (mA cm$^{-2}$) | Overpotential values pre-CA (V vs RHE) | Overpotential values post-CA (V vs RHE) |
| 1 day | IPA/DI | -0.65 | -72.203 | -0.660 ± 0.002 | -0.209 ± 0.010 |
| | NMP | -0.70 | -66.033 | -0.707 ± 0.003 | -0.254 ± 0.003 |
| 1 month | IPA/DI | -0.70 | -54.087 | -0.691 ± 0.004 | -0.292 ± 0.001 |
| | NMP | -0.75 | -25.284 | -0.754 ± 0.004 | -0.327 ± 0.004 |
| 1 year | IPA/DI | -0.75 | -22.646 | -0.731 ± 0.009 | -0.442 ± 0.008 |
| | NMP | -0.80 | -15.152 | -0.789 ± 0.005 | -0.508 ± 0.007 |